\pdfoutput=1
\documentclass{svmult}
\usepackage{graphicx}
\usepackage{amsmath}
\usepackage{amssymb}

\newcommand{\rar}{\rightarrow}

\newcommand{\ttt}{\texttt}

\newcommand{\mca}{\mathcal}
\newcommand{\pms}{\mca{\hat{P}}}

\input{diagxy.tex}

\begin{document}

\title*{The Graph Traversal Pattern}
\author{Marko A. Rodriguez$^1$ \and Peter Neubauer$^2$}

\institute{
	AT\&T Interactive
	\texttt{markorodriguez@attinteractive.com}\\	
	\and
	NeoTechnology
	\texttt{peter.neubauer@neotechnology.com}
}

\maketitle

\begin{abstract}
A graph is a structure composed of a set of vertices (i.e.~nodes, dots) connected to one another by a set of edges (i.e.~links, lines). The concept of a graph has been around since the late 19$^\text{th}$ century, however, only in recent decades has there been a strong resurgence in both theoretical and applied graph research in mathematics, physics, and computer science. In applied computing, since the late 1960s, the interlinked table structure of the relational database has been the predominant information storage and retrieval model. With the growth of graph/network-based data and the need to efficiently process such data, new data management systems have been developed. In contrast to the index-intensive, set-theoretic operations of relational databases, graph databases make use of index-free, local traversals. This article discusses the graph traversal pattern and its use in computing.
\end{abstract}

\section{Introduction}

The first paragraph of any publication on graphs usually contains the iconic $G = (V,E)$ definition of a graph. This definition states that a graph is composed of a set of vertices $V$ and a set of edges $E$. Normally following this definition is the definition of the set $E$. For directed graphs, $E \subseteq (V \times V)$ and for undirected graphs, $E \subseteq \{V \times V\}$. That is, $E$ is a subset of all ordered or unordered permutations of $V$ element pairings. From a purely theoretical standpoint, such definitions are usually sufficient for deriving theorems. However, in applied research, where the graph is required to be embedded in reality, this definition says little about a graph's realization. The structure a graph takes in the real-world determines the efficiency of the operations that are applied to it. It is exactly those efficient graph operations that yield an unconventional problem-solving style. This style of interaction is dubbed the graph traversal pattern and forms the primary point of discussion for this article.\footnote{The term \textit{pattern} refers to data modeling/processing patterns found in computing such as the relational pattern, the map-reduce pattern, etc. In this sense, a pattern is a way of approaching a data-centric problem that usually has benefits in terms of efficiency and/or expressibility.}

\section{The Realization of Graphs}

Relational databases have been around since the late 1960s \cite{rdbms:codd1970} and are todays most predominate data management tool. Relational databases maintain a collection of tables. Each table can be defined by a set of rows and a set of columns. Semantically, rows denote objects and columns denote properties/attributes. Thus, the datum at a particular row/column-entry is the value of the column property for that row object. Usually, a problem domain is modeled over multiple tables in order to avoid data duplication. This process is known as data normalization. In order to unify data in disparate tables, a ``join" is used. A join combines two tables when columns of one table refer to columns of another table. Maintaining these references in a consistent state is known as a referential integrity. This is the classic relational database design which affords them their flexibility \cite{join:mishra1992}.

In stark contrast, graph databases do not store data in disparate tables. Instead there is a single data structure---the graph. Moreover, there is no concept of a ``join" operation as every vertex and edge has a direct reference to its adjacent vertex or edge. The data structure is already ``joined" by the edges that are defined. There are benefits and drawbacks to this model. First, the primary drawback is that its difficult to shard a graph (a difficulty also encountered with relational databases that maintain referential integrity). Sharding is the process of partitioning data across multiple machines in order to scale a system horizontally.\footnote{Sharding is easily solved by other database architectures such as key/value stores \cite{dynamo:decandia2007} and document databases \cite{couchdb:lennon2009}. In such systems, there is no explicit linking between data in different ``collections" (i.e.~documents, key/value pairs). Strict partitions of data make it easier to horizontally scale a database \cite{noshare:stonebraker1986}.}  In a graph, with unconstrained, direct references between vertices and edges, there usually does not exist a clean data partition. Thus, it becomes difficult to scale graph databases beyond the confines of a single machine and at the same time, maintain the speed of a traversal across sharded borders. However, at the expense of this drawback there is a significant advantage: there is a constant time cost for retrieving an adjacent vertex or edge. That is, regardless of the size of the graph as a whole, the cost of a local read operation at a vertex or edge remains constant. This benefit is so important that it creates the primary means by which users interact with graph databases---traversals. Graphs offer a unique vantage point on data, where the solution to a problem is seen as abstractly defined traversals through its vertices and edges.\footnote{The space of graph databases is relatively new. While it is possible to model and process a graph in most any type of database (e.g.~relational databases, key/value stores, document databases), a graph database, in the context of this article, is one that makes use of direct references between adjacent vertices and edges. As such, graph databases are those systems that are optimized for graph traversals. The Neo4j graph database is an example of such a database \cite{neo:larsson2008}.}

\subsection{The Indices of Relational Tables}\label{sec:relational}

Imagine that there is a gremlin who is holding a number between 1 and 100 in memory. Moreover, assume that when guessing the number, the gremlin will only reply by saying whether the guessed number is greater than, less than, or equal to the number in memory. What is the best strategy for determining the number in the fewest guesses? On average, the quickest way to determine the number is to partition the space of guesses into equal size chunks. For example, ask if the number is 50. If the gremlin states that its less than 50, then ask, is the number 25? If greater than 25, then ask, is the number 37? Follow this partition scheme until the number is converged upon. The structure that these guesses form over the sequence from 1 to 100 is a binary search tree. On average, this tree structure is more efficient in time than guessing each number starting from 1 and going to 100. This is ultimately the difference between an index-based search and a linear search. If there were no indices for a set, every element of the set would have to be examined to determine if it has a particular property of interest.\footnote{In a relational database, this process is known as a full table scan.} For $n$ elements, a linear scan of this nature runs in $\mca{O}(n)$. When elements are indexed, there exists two structures---the original set of elements and an index of those elements. Typical indices have the convenient property that searching them takes $\mca{O}(\text{log}_2 n)$. For massive sets, the space that indices take is well worth their weight in time.

\begin{figure}[h!]
	\centering
		\includegraphics[width=0.95\textwidth]{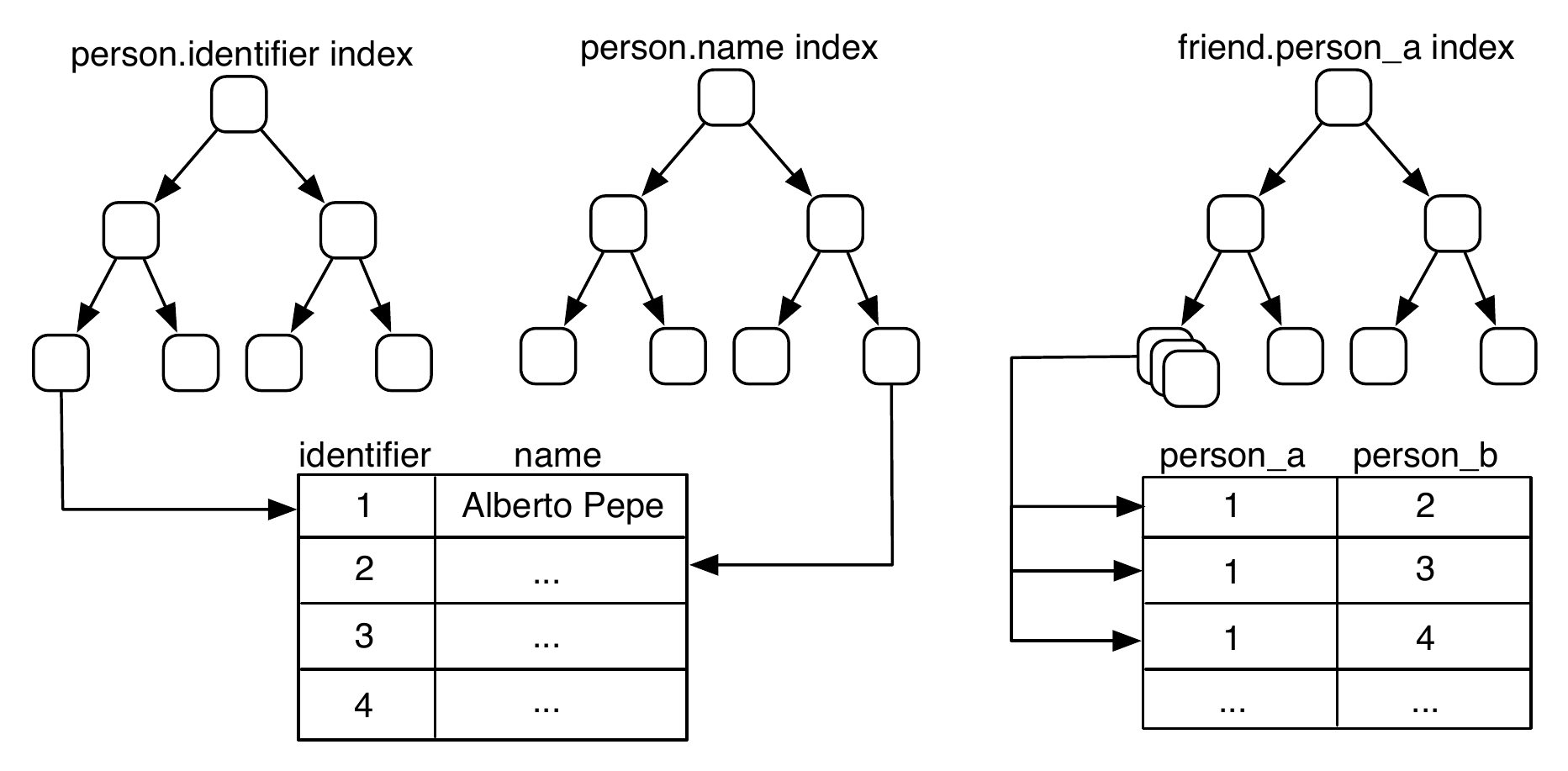}
	\caption{A table representation of people and their friends.}
	\label{fig:friend-table}
\end{figure}

Relational databases take significant advantage of such indices. It is through indices that rows with a column value are efficiently found. Moreover, the index makes it possible to efficiently join tables together in order to move between tables that are linked by particular columns. Assume a simple example where there are two tables: a \ttt{person} table and a \ttt{friend} table. The \ttt{person} table has the following two columns: unique \ttt{identifier} and \ttt{name}. The \ttt{friend} table has the following two columns: \ttt{person\_a} and \ttt{person\_b}. The semantics of the \ttt{friend} table is that person $a$ is friends with person $b$. Suppose the problem of determining the name of all of Alberto Pepe's friends. Figure \ref{fig:friend-table} and the following list breaks down this simple query into all the micro-operations that must occur to yield results.\footnote{Assume that the number of rows in \ttt{person} is $n$ and the number of rows in \ttt{friend} is $m$. Moreover, for the sake of simplicity, assume that names, like identifiers, in the \ttt{person} table are unique.}
\begin{enumerate}
	\item Query the \ttt{person.name} index to find the row in \ttt{person} with the \ttt{name} ``Alberto Pepe." [$\mca{O}(\text{log}_2 n)$]
	\item Given the \ttt{person} row returned by the index, get the \ttt{identifier} for that row.[$\mca{O}(1)$]
	\item Query the \ttt{friend.person\_a} index to find all the rows in \ttt{friend} with the identifier from previous. [$\mca{O}(\text{log}_2 x) : x \ll m$]\footnote{Given that an individual will have many friends, the number of index nodes in the \ttt{friend.person\_a} index will be much less than $m$.}
	\item Given each of the $k$ rows returned, get the \ttt{person\_b} identifier for those rows. [$\mca{O}(k)$]
	\item For each $k$ friend identifiers, query the \ttt{person.identifier} index for the row with friend identifier. [$\mca{O}(k \; \text{log}_2 n)$]
	\item Given the $k$ \ttt{person} rows, get the \ttt{name} value for those rows. [$\mca{O}(k)$]
\end{enumerate}
The final operation yields the names of Alberto's friends. This example elucidates the classic join operation utilized in relational databases. By being able to join the \ttt{person} and \ttt{friend} table, its possible to move from a name, to the person, to his or her friends, and then, ultimately, to their names. In effect, the join operation forms a graph that is dynamically constructed as one table is linked to another table. While having the benefit of being able to dynamically construct graphs, the limitation is that this graph is not explicit in the relational structure, but instead must be inferred through a series of index-intensive operations. Moreover, while only a particular subset of the data in the database may be desired (e.g.~only Alberto's friend's), all data in all queried tables must be examined in order to extract the desired subset (e.g.~all friends of all people). Even though a $\mca{O}(\text{log}_2 n)$ read-time is fast for a search, as the the indices grow larger with the growth of the data and as more join operations are used, this model becomes inefficient. At the limit, the inferred graph that is constructed through joins is best solved (with respects to time), by a graph database.

\subsection{The Graph as an Index}\label{sec:graph}

Most of graph theory is concerned with the development of theorems for single-relational graphs \cite{graphs:chartrand1986}. A single-relational graph maintains a set of edges, where all the edges are homogeneous in meaning. For example, all edges denote friendship or kinship, but not both together within the same structure. In application, complex domain models are more conveniently represented by multi-relational, property graphs.\footnote{In the parlance of graphs, a property graph is a directed, edge-labeled, attributed multi-graph. For the sake of simplicity, such structures will simply be called property graphs. These types of graph structures are used extensively in computing as they are more expressive than the simplified mathematical objects studied in theory. However, note that expressiveness is defined by ease of use, not by the limits of what can be modeled \cite{mapnetwork:rodriguez2009}.} The edges in a property graph are typed or labeled and thus, edges are heterogenous in meaning. For example, a property graph can model friendship, kinship, business, communication, etc. relationships all within the same structure. Moreover, vertices and edges in a property graph maintain a set of key/value pairs. These are known as properties and allow for the representation of non-graphical data---e.g.~the name of a vertex, the weight of an edge, etc. Formally, a property graph can be defined as $G = (V,E,\lambda,\mu)$, where edges are directed (i.e.~$E \subseteq (V \times V)$), edges are labeled (i.e.~$\lambda: E \rar \Sigma$), and properties are a map from elements and keys to values (i.e.~$\mu: (V \cup E) \times R \rar S$). 

In the property graph model, it is common for the properties of the vertices (and sometimes edges) to be indexed using a tree structure analogous, in many ways, to those used by relational databases. This index can be represented by some external indexing system or endogenous to the graph as an embedded tree (see \S \ref{sec:endogenous}).\footnote{The reason for using an external indexing system is that it may be optimized for certain types of lookups such as full-text search.} Given the prior situation, once a set of elements have been identified by the index search, then a traversal is executed through the graph.\footnote{This is ultimately what is accomplished in a relational database when a row of a table is located and a value in a column of that row is fetched (e.g.~see the second micro-operation of the relational database enumeration previous). However, when that row doesn't have all the requisite data (usually do to database normalization), it requires the joining with another table to locate that data. It is this situation which is costly in a relational database.} Elements in a graph are adjacent to one another by direct references. A vertex is adjacent to its incoming and outgoing edges and an edge is adjacent to its outgoing (i.e.~tail) and incoming (i.e.~head) vertices. The domain model defines how the elements of the problem space are related. Similar to the gremlin stating that 50 is greater than the number to be guessed, an edge connecting vertex $i$ and $j$ and labeled \ttt{friend} states that vertex $i$ is \ttt{friend} related to vertex $j$. Indices create ``short cuts" in the graph as they partition elements according to specialized, compute-centric semantics (e.g.~numbers being less than or greater than another). Likewise, a domain model partitions elements using semantics defined by the domain modeler. Thus, in many ways, a graph can be seen as an indexing structure.

\begin{figure}[h!]
	\centering
		\includegraphics[width=0.85\textwidth]{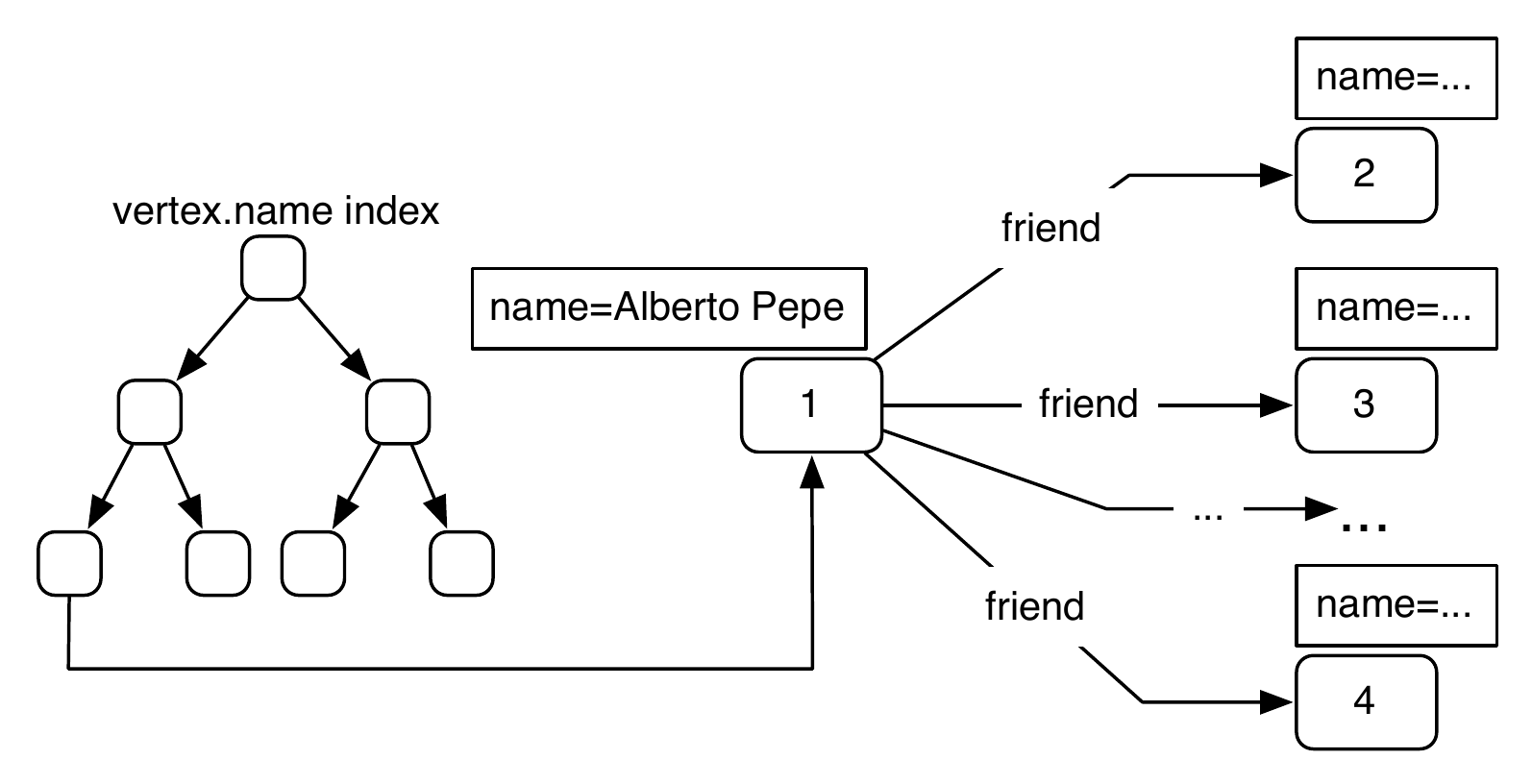}
	\caption{A graph representation of people and their friends. Given the tree-nature of the \ttt{vertex.name} index, it is possible, and many times useful to model the index endogenous to the graph (see \S \ref{sec:endogenous}).}
	\label{fig:friend-graph}
\end{figure}

In the relational example previous, a person in the \ttt{person} table has two properties: a unique \ttt{identifier} and a \ttt{name}. The analogue in a property graph would be to have the \ttt{identifier} and \ttt{name} values represented as vertex properties. Moreover, the \ttt{friend} table would not exist as a table, but as direct \ttt{friend}-labeled edges between vertices. This idea is diagrammed in Figure \ref{fig:friend-graph}. The micro-operations used to find the name of all of Alberto Pepe's friends are provided in the following enumeration.
\begin{enumerate}
	\item Query the \ttt{vertex.name} index to find all the vertices in $G$ with the name ``Alberto Pepe." [$\mca{O}(\text{log}_2 n)$]
	\item Given the vertex returned, get the $k$ \ttt{friend} edges emanating from this vertex. [$\mca{O}(k+x)$]\footnote{If a graph database does not index the edges of a vertex by their labels, then a linear scan of all edges emanating from a vertex must occur to locate the set of \ttt{friend}-labeled edges. Thus, $k+x$ is the total number of edges emanating from the current vertex.}
	\item Given the $k$ \ttt{friend} edges retrieved, get the $k$ vertices on the heads of those edges. [$\mca{O}(k)$]
	\item Given these $k$ vertices, get the $k$ \ttt{name} properties of these vertices. [$\mca{O}(ky)$]\footnote{If a graph database does not index the properties of a vertex, then a linear scan of all the properties must occur. If $y$ is the total number of properties on the vertices (assuming a homogenous count for all vertices), then, in the worst case scenario, $ky$ elements must be examined.} 
\end{enumerate}
The final operation yields the names of Alberto's friends. In a graph database, there is no explicit join operation because vertices maintain direct references to their adjacent edges. In many ways, the edges of the graph serve as explicit, ``hard-wired" join structures (i.e.~structures that are not computed at query time as in a relational database). The act of traversing over an edge is the act of joining. However, what makes this more efficient in a graph database is that traversing from one vertex to another is a constant time operation. Thus, traversal time is defined solely by the number of elements touched by the traversal. This is irrespective of the size/topology of the graph as a whole. The time it takes to make a single step in a traversal is determined by the local topology of the subgraph surrounding the particular vertex being traversed from.\footnote{The consequence of this is that traversing through a ``super node" (i.e.~a high-degree vertex) in a graph is slower than traversing through a small-degree vertex.}

The real power of graph databases makes itself apparent when traversing multiple steps in order to unite disparate vertices by a path (i.e.~vertices not directly connected). First, there are no $\mca{O}(\text{log}_2 n)$ operations. Second, the type of path taken, defines the ``higher order," inferred relationship that exists between two vertices.\footnote{In many ways, this is the graph equivalent of the join operation used by relational databases---though no global indices are used. When traversing a multi-step path, the source and sink vertices are united by a semantic determined by the path taken. For example, going from a person, to their friends, and then to their friends friends, will unite that person to people two-steps away in the graph. This popular path is known FOAF (friend of a friend).} Traversals based on abstractly defined paths is the core of the graph traversal pattern. The next section discusses the graph traversal pattern and its application to common problem-solving situations.

\section{Graph Traversals}

A traversal refers to visiting elements (i.e.~vertices and edges) in a graph in some algorithmic fashion.\footnote{In general, the term ``algorithm" is used in a looser sense than the classic definition in that it allows for randomization and sampling when traversing.} This section will present a functional, flow-based approach \cite{flow:morrison1994} to traversing property graphs and how different types of traversals over different types of graph datasets support different types of problem-solving.

The most primitive, read-based operation on a graph is a single step traversal from element $i$ to element $j$, where $i,j \in (V \cup E)$.\footnote{While it is possible to write and delete elements from a graph, such operations will not be discussed.} For example, a single step operation can answer questions such as ``which edges are outgoing from this vertex?", ``which vertex is at the head of this edge?", etc. Single step operations expose explicit adjacencies in the graph (i.e.~adjacencies that are ``hard-wired"). The following list itemizes the various types of single step traversals. Note that these operations are defined over power multiset domains and ranges.\footnote{The power set of set $A$ is denoted $\mca{P}(A)$ and is the set of all subsets of $A$ (i.e.~$2^{A}$). The power multiset of $A$, denoted $\pms(A)$, is the infinite set of all subsets of multisets of $A$. This set is infinite because multisets allow for repeated elements \cite{multiset:monro1987}.} The reason for this is that is naturally allows for function composition, where a composition is a formal description of a traversal.\footnote{The path algebra defined in \cite{pathalg:rodriguez2009} operates over multi-relational graphs represented as a tensor. Besides the inclusion of vertex/edge properties used in this article, the tensor-based path algebra has the same expressivity as the functional model presented in this section.}
\begin{itemize}
	\item $e_\text{out}: \pms(V) \rar \pms(E)$: traverse to the outgoing edges of the vertices.
	\item $e_\text{in}: \pms(V) \rar \pms(E)$: traverse to the incoming edges to the vertices.
	\item $v_\text{out}: \pms(E) \rar \pms(V)$: traverse to the outgoing (i.e.~tail) vertices of the edges.
	\item $v_\text{in}: \pms(E) \rar \pms(V)$: traverse the incoming (i.e.~head) vertices of the edges.
	\item $\epsilon: \pms(V \cup E) \times R \rar \pms(S)$: get the element property values for key $r \in R$.
\end{itemize}
When edges are labeled and elements have properties, it is desirable to constrain the traversal to edges of a particular label or elements with particular properties. These operations are known as filters and are abstractly defined in the following itemization.\footnote{Filters can be defined as allowing or disallowing certain elements. For allowing, the symbol $+$ is used. For disallowing, the symbol $-$ is used.}
\begin{itemize}
	\item $e_{\text{lab}\pm}: \pms(E) \times \Sigma \rar \pms(E)$: allow (or filter) all edges with the label $\sigma \in \Sigma$.
	\item $\epsilon_{\text{p}\pm}: \pms(V \cup E) \times R \times S \rar \pms(V \cup E)$: allow (or filter) all elements with the property $s \in S$ for key $r \in R$.
	\item $\epsilon_{\epsilon\pm}: \pms(V \cup E) \times (V \cup E) \rar \pms(V \cup E)$: allow (or filter) all elements that are the provided element.
\end{itemize}

Through function composition, we can define graph traversals of arbitrary length. A simple example is traversing to the names of Alberto Pepe's friends. If $i$ is the vertex representing Alberto Pepe and
\begin{equation*}
f: \pms(V) \rar \pms(S),
\end{equation*}
where
\begin{equation*}
f(i) = \epsilon\left(v_\text{in}\left(e_\text{lab+}\left(e_\text{out}(i), \text{friend}\right)\right),\text{name}\right),
\end{equation*}
then $f(i)$ will return the names of Alberto Pepe's friends. Through function currying and composition, the previous definition can be represented more clearly with the following function rule,
\begin{equation*}
f(i) = \left(\epsilon^\text{name} \circ v_\text{in} \circ e_\text{lab+}^\text{friend} \circ e_\text{out}\right)(i).
\end{equation*}
\begin{figure}[h!]
	\centering
		\includegraphics[width=0.6\textwidth]{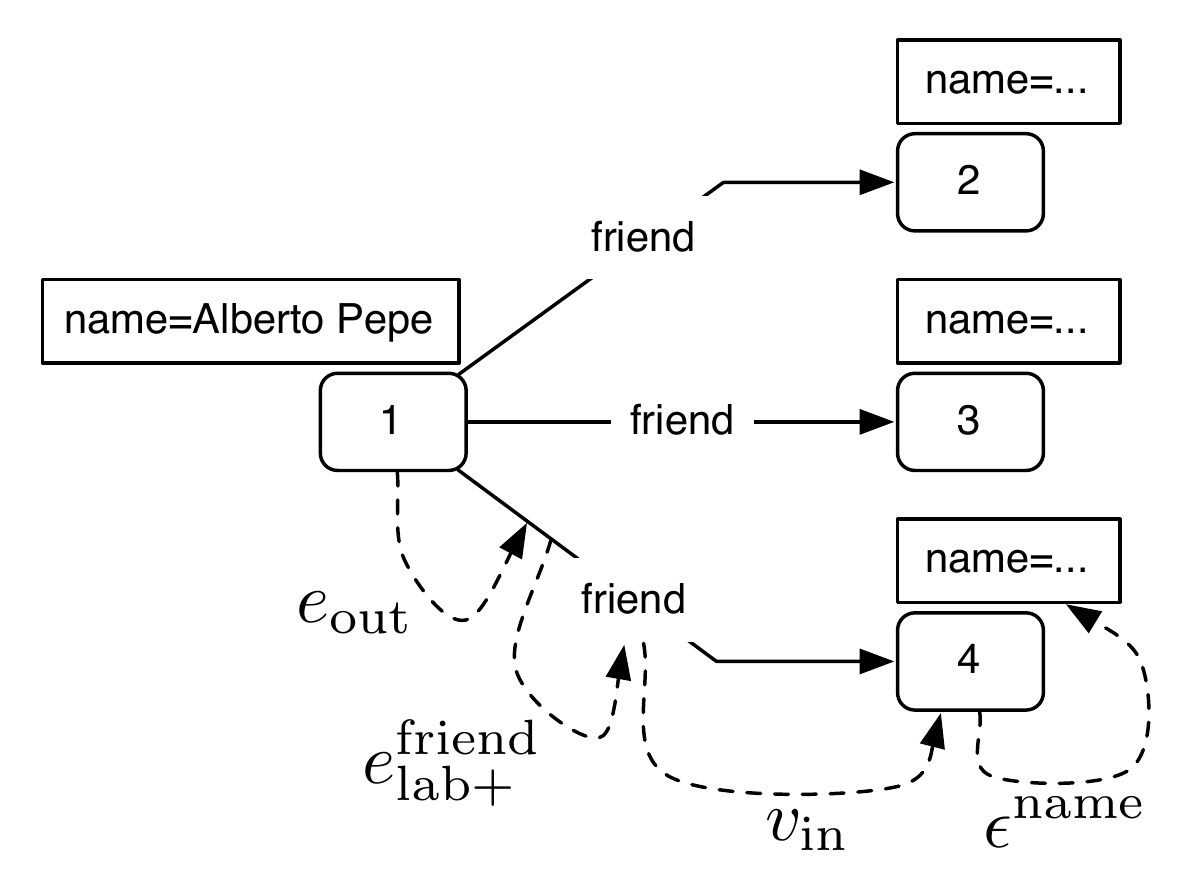}
	\caption{A single path along along the $f$ traversal.}
	\label{fig:friend-graph-names}
\end{figure}
The function $f$ says, traverse to the outgoing edges of vertex $i$, then only allow those edges with the label \ttt{friend}, then traverse to the incoming (i.e.~head) vertices on those \ttt{friend}-labeled edges. Finally, of those vertices, return their \ttt{name} property.\footnote{Note that the order of a composition is evaluated from right to left.} A single legal path according to this function is diagrammed in Figure \ref{fig:friend-graph-names}. Though not diagrammed for the sake of clarity, the traversal would also go from vertex $1$ to the name of vertex $2$ and vertex $3$. The function $f$ is a ``higher-order" adjacency defined as the composition of explicit adjacencies and serves as a join of Alberto and his friend's names.\footnote{This is known as a virtual edge in the graph system called DEX \cite{dex:martinez2007}.} The remainder of this section demonstrates graph traversals in real-world problems-solving situations.

\subsection{Traversing for Recommendation}\label{sec:recommendation}

\begin{figure}[h!]
	\centering
		\includegraphics[width=0.7\textwidth]{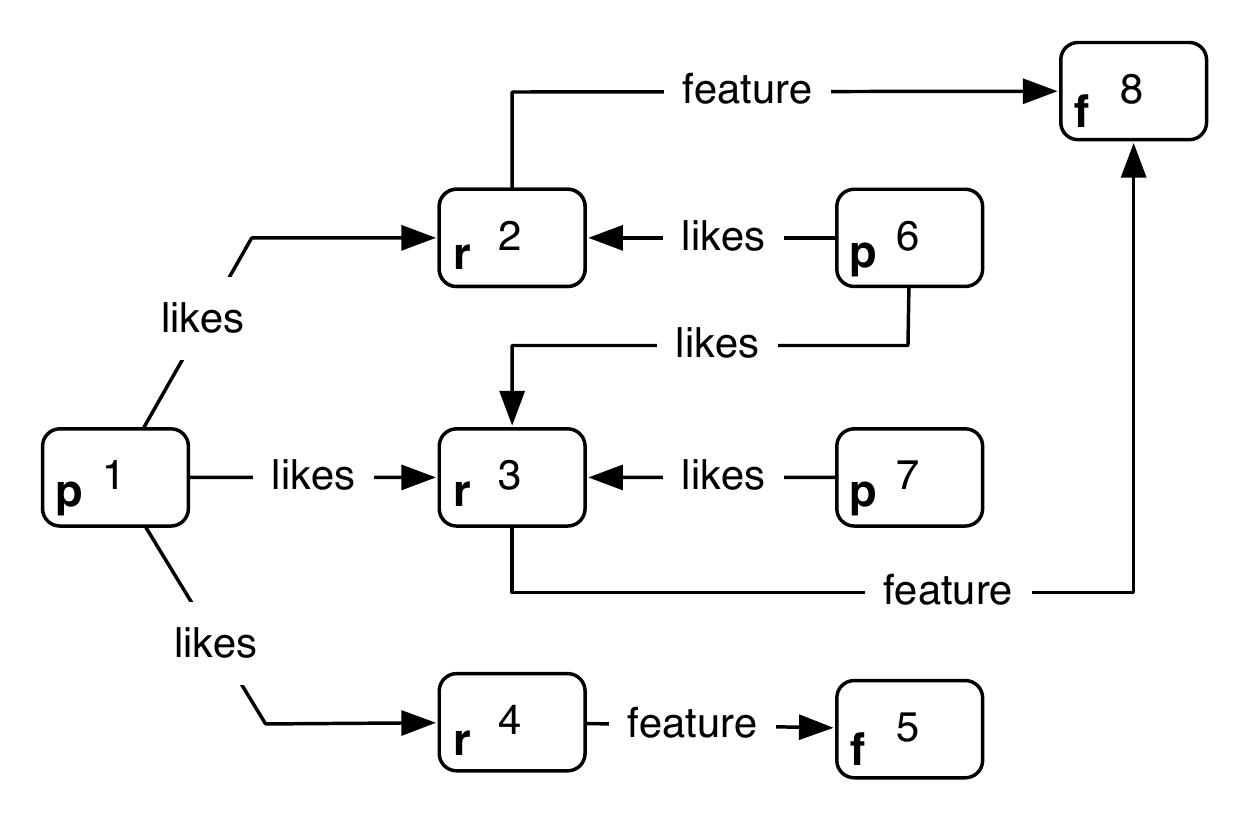}
	\caption{A graph data structure containing people (p), their liked resources (r), and each resource's features (f).}
	\label{fig:rec-graph}
\end{figure}

Recommendation systems are designed to help people deal with the problem of information overload by filtering information in the system that doesn't pertain to the person \cite{rec:perugini2004}. In a positive sense, recommendation systems focus a person's attention on those resources that are likely to be most relevant to their particular situation. There is a standard dichotomy in recommendation research---that of content- vs. collaborative filtering-based recommendation. The prior deals with recommending resources that share characteristics (i.e.~content) with a set of resources. The latter is concerned with determining the similarity of resources based upon the similarity of the taste of the people modeled within the system \cite{collab:herlocker2006}. These two seemingly different techniques to recommendation are conveniently solved using a graph database and two simple traversal techniques \cite{rec:mirza2003,griffen:spread2006}. Figure \ref{fig:rec-graph} presents a toy graph data set, where there exist a set of people, resources, and features related to each other by \ttt{likes}- and \ttt{feature}-labeled edges. This simple data set is used for the remaining examples of this subsection.

\subsubsection{Content-Based Recommendation}

\begin{figure}[h!]
	\centering
		\includegraphics[width=0.525\textwidth]{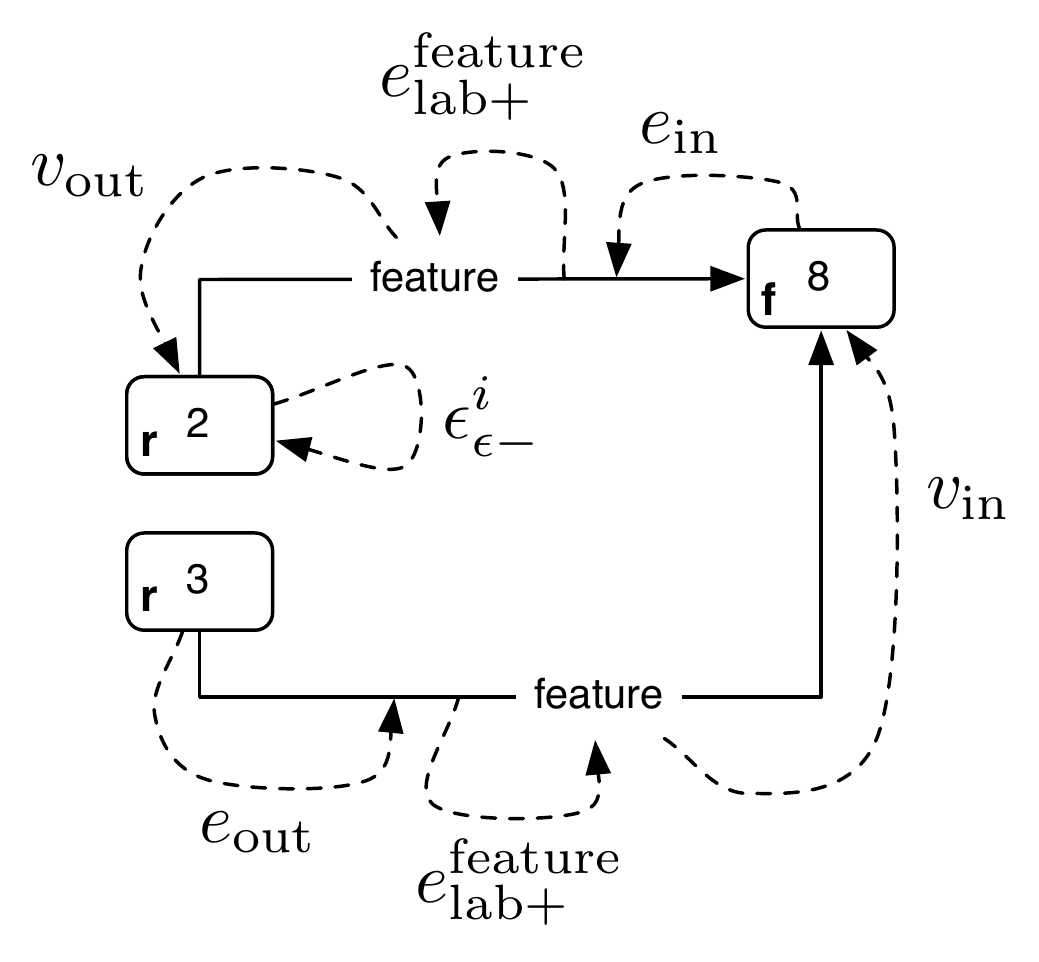}
	\caption{A traversal that identifies resources that are similar in content to a set of resources based upon shared features.}
	\label{fig:rec-graph-content}
\end{figure}

In order to identify resources that that are similar in features (i.e.~content-based recommendation) to a resource, traverse to all resources that share the same features. This is accomplished with the following function, $f : \pms(V) \rar \pms(V)$, where
\begin{equation*}
f(i) =  \left(\epsilon_{\epsilon-}^i \circ v_\text{out} \circ e_\text{lab+}^\text{feature} \circ e_\text{in} \circ v_\text{in} \circ e_\text{lab+}^\text{feature} \circ e_\text{out}\right)(i).
\end{equation*}
Assuming $i = 3$, function $f$ states, traverse to the outgoing edges of resource vertex $3$, only allow \ttt{feature}-labeled edges, and then traverse to the incoming vertices of those \ttt{feature}-labeled edges. At this point, the traverser is at feature vertex $8$. Next, traverse to the incoming edges of feature vertex $8$, only allow \ttt{feature}-labeled edges, and then traverse to the outgoing vertices of these \ttt{feature}-labeled edges. At this point, the traverser is at resource vertices $3$ and $2$. However, since we are trying to identify those resources similar in content to vertex $3$, we need to filter out vertex $3$. This is accomplished by the last stage of the function composition. Thus, given the toy graph data set, vertex $2$ is similar to vertex $3$ in content. This traversal is diagrammed in Figure \ref{fig:rec-graph-content}.

Its simple to extend content-based recommendation to problems such as: ``Given what person $i$ likes, what other resources have similar features?" Such a problem is solved using the previous function $f$ defined above combined with a new composition that finds all the resources that person $i$ likes. Thus, if $g : \pms(V) \rar \pms(V)$, where
\begin{equation*}
g(i) = \left(v_\text{in} \circ e_\text{lab+}^\text{likes} \circ e_\text{out}\right)(i),
\end{equation*}
then to determine those resources similar in features to the resources that person vertex $7$ likes, compose function $f$ and $g$: $(f \circ g)(7)$. Those resources that share more features in common will be returned more by $f \circ g$.\footnote{Again, path traversal functions are defined over power multisets. In this way, its possible for a function to return repeated elements. In some situations, deduplicating this set is desired. In other situations, repeated elements can be used to weight/rank the results.} 

What has been presented is an example of the use of traversals to do na\"ive content-based recommendation. It is possible to extend the functions presented to normalize paths (e.g.~a resource can have every feature and thus, is related to everything), find novelty (e.g.~feature paths that are rare and only shared by a certain subset of resources), etc. In most cases, when creating a graph traversal, a developer will compose different predefined paths into a longer compositions. Along with speed of execution, this is one of the benefits of using a functional, flow-based model for graph traversals \cite{graphflow:yoo2009}. Moreover, each component has a high-level meaning (e.g.~the resources that a person likes) and as such, the verbosity of longer compositions can be minimal (e.g.~$f \circ g$).

\subsubsection{Collaborative Filtering-Based Recommendation}

\begin{figure}[h!]
	\centering
		\includegraphics[width=0.7\textwidth]{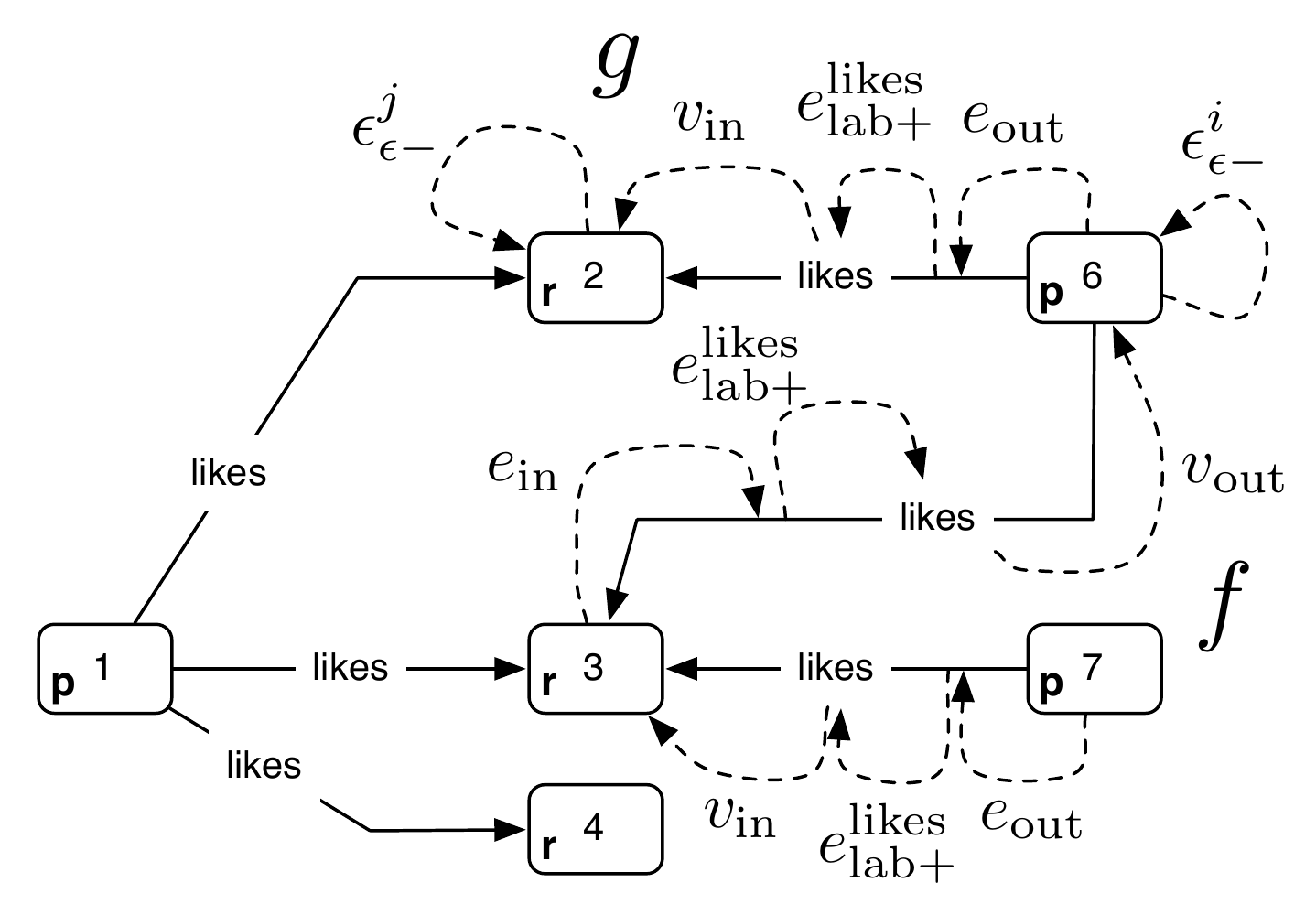}
	\caption{A traversal that identifies resources that are similar in content to a resource based upon shared features.}
	\label{fig:rec-graph-collab}
\end{figure}

With collaborative filtering, the objective is to identify a set of resources that have a high probability of being liked by a person based upon identifying other people in the system that enjoy similar likes. For example, if person $a$ and person $b$ share 90\% of their liked resources in common, then the remaining 10\% they don't share in common are candidates for recommendation. Solving the problem of collaborative filtering using graph traversals can be accomplished with the following traversal. For the sake of clarity, the traversal is broken into two components: $f$ and $g$, where $f : \pms(V) \rar \pms(V)$ and $g : \pms(V) \rar \pms(V)$.
\begin{equation*}
f(i) =  \left(\epsilon_{\epsilon-}^i \circ v_\text{out} \circ e_\text{lab+}^\text{like} \circ e_\text{in} \circ v_\text{in} \circ e_\text{lab+}^\text{like} \circ e_\text{out}\right)(i).
\end{equation*}
Function $f$ traverses to all those people vertices that like the same resources as person vertex $i$ and who themselves are not vertex $i$ (as a person is obviously similar to themselves and thus, doesn't contribute anything to the computation). The more resources liked that a person shares in common with $i$, the more traversers will be located at that person's vertex. In other words, if person $i$ and person $j$ share $10$ liked resources in common, then $f(i)$ will return person $j$ $10$ times. Next, function $g$ is defined as
\begin{equation*}
g(j) =  \left(v_\text{in} \circ e_\text{lab+}^\text{like} \circ e_\text{out}\right)(j).
\end{equation*}
Function $g$ traverses to all the resources liked by vertex $j$. In composition, $(g \circ f)(i)$ determines all those resources that are liked by those people that have similar tastes to vertex $i$. If person $j$ likes $10$ resources in common with person $i$, then the resources that person $j$ likes will be returned at least $10$ times by $g \circ f$ (perhaps more if a path exists to those resources from another person vertex as well). Figure \ref{fig:rec-graph-collab} diagrams a function path starting from vertex $7$. Only one legal path is presented for the sake of diagram clarity.

With the graph traversal pattern, there exists a single graph data structure that can be traversed in different ways to expose different types of recommendations---generally, different types of relationships between vertices. Being able to mix and match the types of traversals executed alters the semantics of the final rankings and conveniently allows for hybrid recommendation algorithms to emerge.

\subsection{Traversing Endogenous Indices}\label{sec:endogenous}

A graph is a general-purpose data structure. A graph can be used to model lists, maps, trees, etc. As such, a graph can model an index. It was assumed, in \S \ref{sec:graph}, that a graph database makes use of an external indexing system to index the properties of its vertices and edges. The reason stated was that specialized indexing systems are better suited for special-purpose queries such as those involving full-text search. However, in many cases, there is nothing that prevents the representation of an index within the graph itself---vertices and edges can be indexed by other vertices and edges.\footnote{One of the primary motivations behind this article is to stress the importance of thinking of a graph as simply an index of itself, where the primary purpose is to traverse the various defined indices in ways that elicit problem-solving within the domain being modeled.} In fact, given the nature of how vertices and edges directly reference each other in a graph database, index look-up speeds are comparable. Endogenous indices afford graph databases a great flexibility in modeling a domain. Not only can objects and their relationships be modeled (e.g.~people and their friendships), but also the indices that partition the objects into meaningful subsets (e.g.~people within a 2D region of space).\footnote{Those indices that have a graph-like structure are suited for representing as a graph. It is noted that not all indices meet this criteria.} The remainder of this subsection will discuss the representation and traversal of a spatial, 2D-index that is explicitly modeled within a property graph.

\begin{figure}[h!]
	\centering
		\includegraphics[width=0.35\textwidth]{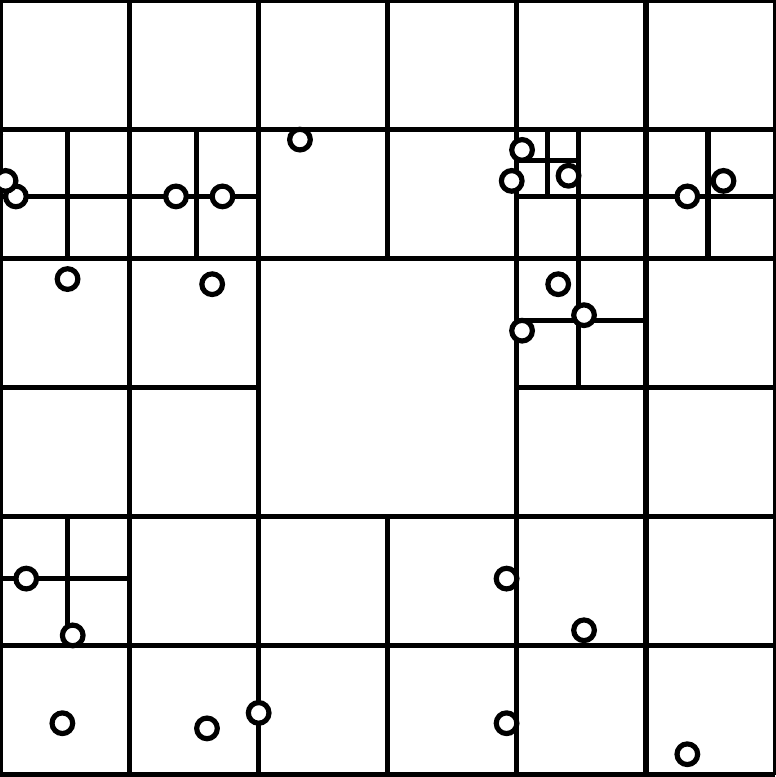}
	\caption{A quadtree partition of a plane. This figure is an adaptation of a public domain image provided courtesy of David Eppstein.}
	\label{fig:quadtree-partition}
\end{figure}

The domain of spatial analysis makes use of advanced indexing structures such as the quadtree \cite{quadtree:finkel1974,multi:samet2006}. Quadtrees partition a two-dimensional plane into rectangular boxes based upon the spatial density of the points being indexed. Figure \ref{fig:quadtree-partition} diagrams how space is partitioned as the density of points increases within a region of the index.

\begin{figure}[h!]
	\centering
		\includegraphics[width=0.8\textwidth]{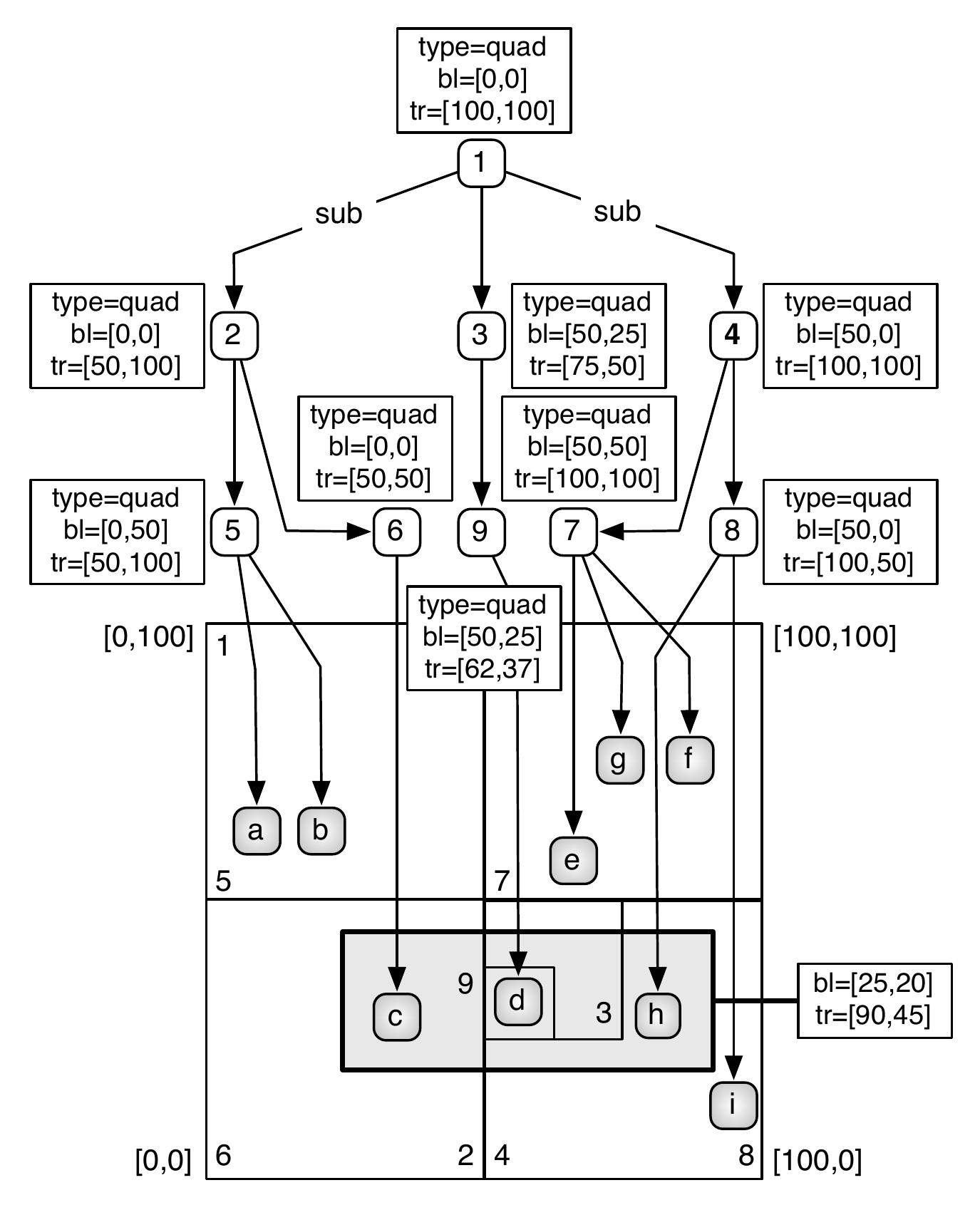}
	\caption{A quadtree index of a space that contains points of interest. The index is composed of the vertices $1$-$9$ and the points of interest are the vertices $a$-$i$. While not diagrammed for the sake of clarity, all edges are labeled \ttt{sub} (meaning subsumes) and each point of interest vertex has an associated bottom-left (bl) property, top-right (tr) property, and a type property which is equal to ``poi."}
	\label{fig:quadtree}
\end{figure}

In order to demonstrate how a quadtree index can be represented and traversed, a toy graph data set is presented. This data set is diagrammed in Figure \ref{fig:quadtree}. 
The top half of Figure \ref{fig:quadtree} represents a quadtree index (vertices $1$-$9$). This quadtree index is partitioning ``points of interest" (vertices $a$-$i$) located within the diagrammed plane.\footnote{The plane depicted does not actually exist as a data structure, but is represented here to denote how the different vertices lying on that plane are spatially located (i.e.~spatial information is represented explicitly in the properties of the vertices). Thus, vertices closer to each other on the plane are closer together.} All vertices maintain three properties---bottom-left (bl), top-right (tr), and type. For a quadtree vertex, these properties identify the two corner points defining a rectangular bounding box (i.e.~the region that the quadtree vertex is indexing) and the vertex type which is equal to ``quad". For a point of interest vertex, these properties denote the region of space that the point of interest exists within and the vertex type which is equal to ``poi."

Quadtree vertex $1$ denotes the entire region of space being indexed. This region is defined by its bottom-left (bl) and top-right (tr) corner points---namely $[0,0]$ and $[100,100]$, where $\text{bl}_x = 0$, $\text{bl}_y = 0$, $\text{tr}_x = 100$, and $\text{tr}_y = 100$. Within the region defined by vertex $1$, there are $8$ other defined regions that partition that space into smaller spaces (vertices $2$-$9$). When one vertex subsumes another vertex by a directed edge labeled \ttt{sub} (i.e.~subsumes), the outgoing (i.e.~tail) vertex is subsuming the space that is defined by the incoming (i.e.~head) vertex. Given these properties and edges, identifying point of interest vertices within a region of space is simply a matter of traversing the quadtree index in a directed/algorithmic fashion.

In Figure \ref{fig:quadtree}, the shaded region represents the spatial query: ``Which points of interest are within the rectangular region defined by the corner points $\text{bl}=[25,20]$ and $\text{tr}=[90,45]$?" In order to locate all the points of interest in this region, iteratively execute the following traversal starting from the root of the quadtree index (i.e.~vertex $1$). The function is defined as $f : \pms(V) \rar \pms(V)$, where
\begin{equation*}
f(i) = \left(\epsilon_{\text{p+}}^{\text{tr}_y \geq 20} \circ \epsilon_{\text{p+}}^{\text{tr}_x \geq 25} \circ \epsilon_{\text{p+}}^{\text{bl}_y \leq 45} \circ \epsilon_{\text{p+}}^{\text{bl}_x \leq 90} \circ v_\text{in} \circ e_\text{lab+}^\text{sub} \circ e_\text{out}\right)(i).
\end{equation*}
The defining aspect of $f$ is the set of 4 $\epsilon_\text{p+}$ filters that determine whether the current vertex is overlapping or within the query rectangle. Those vertices not overlapping or within the query rectangle are not traversed to. Thus, as the traversal iterates, fewer and fewer paths are examined and the resulting point of interest vertices within the query rectangle are converged upon. With respect to Figure \ref{fig:quadtree}, after 3 iterations of $f$, the traversal will have returned all the points of interest within the query rectangle. The first iteration, will traverse to the index vertices 2, 3, and 4. The second iteration will traverse to the vertices 6, 8 and 9. Note that vertices 5 and 7 do not meet the criteria of the $\epsilon_\text{p+}$ filters. Finally, on the third iteration, the traversal returns vertices $c$, $d$, and $h$. Note that vertex $i$ is not returned because it, like 5 and 7, does not meet the $\epsilon_\text{p+}$ filter criteria. A summary of the legal vertices traversed to at each iteration is enumerate below.
\begin{enumerate}
	\item 2, 3, 4
	\item 6, 9, 8
	\item $c$, $d$, $h$
\end{enumerate}

There is a more efficient traversal that can be evaluated. If the bounding box defined by a quadtree vertex is completely subsumed by the query rectangle (i.e.~not just overlapping), then, at that branch in the traversal, the traverser no longer needs to evaluate the $\epsilon_{\text{p+}}$-region filters and, as such, can simply iterate all the way down \ttt{sub}-labeled edges to the point of interest vertices knowing that they are completely within the query rectangle. For example, in Figure \ref{fig:quadtree}, once it is realized that vertex 9 is completely within the query rectangle, then the location properties of vertex $d$ do not need to be examined.\footnote{In general, disregarding bounding box property checks holds for both quadtree vertices and point of interest vertices that are subsumed by a quadtree vertex that is completely within the query rectangle.} The functions that define this traversal and the composition of these functions into a flow graph is defined below, where $A \subset \pms(V)$ is the multiset of all quadtree index vertices overlapping or within the query rectangle, $B \subseteq A$ is the multiset of all quadtree index vertices completely within the query rectangle, and $C \subset \pms(V)$ is the multiset of all point of interest vertices overlapping or within the query rectangle.
\begin{eqnarray*}
f(i) &=& \left(\epsilon_{\text{p+}}^{\text{tr}_y \geq 20} \circ \epsilon_{\text{p+}}^{\text{tr}_x \geq 25} \circ \epsilon_{\text{p+}}^{\text{bl}_y \leq 45} \circ \epsilon_{\text{p+}}^{\text{bl}_x \leq 90} \circ \epsilon_{\text{p+}}^{\text{type} = \text{quad}} \circ v_\text{in} \circ e_\text{lab+}^\text{sub} \circ e_\text{out}\right)(i) \\
g(i) &=& \left(\epsilon_{\text{p+}}^{\text{tr}_y \leq 45} \circ \epsilon_{\text{p+}}^{\text{tr}_x \leq 90} \circ \epsilon_{\text{p+}}^{\text{bl}_y \geq 20} \circ \epsilon_{\text{p+}}^{\text{bl}_x \geq 25} \right)(i) \\
h(i) &=& \left(\epsilon_{\text{p+}}^{\text{type} = \text{quad}} \circ v_\text{in} \circ e_\text{lab+}^\text{sub} \circ e_\text{out}\right)(i) \\ 
s(i) &=& \left(\epsilon_{\text{p+}}^{\text{tr}_y \geq 20} \circ \epsilon_{\text{p+}}^{\text{tr}_x \geq 25} \circ \epsilon_{\text{p+}}^{\text{bl}_y \leq 45} \circ \epsilon_{\text{p+}}^{\text{bl}_x \leq 90} \circ \epsilon_{\text{p+}}^{\text{type} = \text{poi}} \circ v_\text{in} \circ e_\text{lab+}^\text{sub} \circ e_\text{out}\right)(i) \\
r(i) &=& \left(\epsilon_{\text{p+}}^{\text{type} = \text{poi}} \circ v_\text{in} \circ e_\text{lab+}^\text{sub} \circ e_\text{out}\right)(i) \\ 
\end{eqnarray*}
\scalefactor{0.85}
\begin{equation*}
\bfig
\Vtriangle(0,0)[\;\;\;`\;\;\;`C;g`s`r]
\Loop(0,500)A(ur,ul)_f
\Loop(1000,500)B(ur,ul)_h
\efig
 \end{equation*}
Function $f$ traverses to those quadtree vertices that overlap or are within the query rectangle. Function $g$ allows only those quadtree vertices that are completely within the query rectangle. Function $h$ traverses to subsumed quadtree vertices. Function $s$ traverses to point of interest vertices that are overlapping or within the query rectangle. Finally, function $r$ traverses to subsumed point of interest vertices. Note that functions $h$ and $r$ do no check the bounding box properties of their domain vertices. As a quadtree becomes large, this becomes a more efficient solution to finding all points of interest within a query rectangle.

The ability to model an index endogenous to a graph allows the domain modeler to represent not only objects and their relations (e.g.~people and their friendships), but also ``meta-objects" and their relationships (e.g.~index nodes and their subsumptions). In this way, the domain modeler can organize their model according to partitions that make sense to how the model will be used to solve problems. Moreover, by combining the traversal of an index with the traversal of a domain, there exists a single unified means by which problems are solved within a graph database---the graph traversal pattern.

\section{Conclusion}

Graphs are a flexible modeling construct that can be used to model a domain and the indices that partition that domain into an efficient, searchable space. When the relations between the objects of the domain are seen as vertex partitions, then a graph is simply an index that relates vertices to vertices by edges. The way in which these vertices relate to each other determines which graph traversals are most efficient to execute and which problems can be solved by the graph data structure. Graph databases and the graph traversal pattern do not require a global analysis of data. For many problems, only local subsets of the graph need to be traversed to yield a solution. By structuring the graph in such a way as to minimize traversal steps, limit the use of external indices, and reduce the number of set-based operations, modelers gain great efficiency that is difficult to accomplish with other data management solutions.

\section*{Acknowledgements} 

This article was made possible by the Center for Nonlinear Studies of the Los Alamos National Laboratory, AT\&T Interactive, NeoTechnology, and the help and support of Tobias Ivarsson, Craig Taverner, Johan Svensson, and Jennifer H. Watkins. This work was partially funded by the Los Alamos National Laboratory grant LDRD:20080724PRD2 (entitled ``Towards Human Level Artificial Intelligence: A Cortically Inspired Semantic Network Approach to Information Processing and Storage").

\end{document}